\documentclass[useAMS,usenatbib]{mn2e}

\usepackage{graphicx, amssymb,amsmath}

\title[A cold stream around NGC1851?]{Spectroscopic hint of a cold stream in the direction of the globular cluster NGC
1851\thanks{Based on VIMOS observations collected with the Very Large
Telescope at the European Southern Observatory, Cerro Paranal,
Chile, within the observing program 082.D-0244}}
\author[Sollima et al.]{A. Sollima$^{1}$\thanks{E-mail:
antonio.sollima@oapd.inaf.it}, R. G. Gratton$^{1}$, J. A. Carballo-Bello$^{2,3}$, 
D. Mart{\'{\i}}nez-Delgado$^{4,5}$,\\ 
\newauthor
E. Carretta$^{6}$, A. Bragaglia$^{6}$, S. Lucatello$^{1}$, J. Pe{\~n}arrubia$^{7}$\\
$^{1}$ INAF Osservatorio Astronomico di Padova, vicolo dell'Osservatorio 5,
35122, Padova, Italy\\
$^{2}$ Instituto de Astrofisica de Canarias, Calle via Lactea s/n, 38205, La Laguna, Spain\\
$^{3}$ Universidad de La Laguna, Departamento de Astrofisica, 38205, La Laguna, Spain\\
$^{4}$ Max-Planck Institut fur Astronomie, Konigstuhl 17, 69117, Heidelberg,
Germany\\
$^{4}$ Alexander von Humboldt Fellow for Advanced Research\\
$^{6}$ INAF Osservatorio Astronomico di Bologna, via Ranzani 1, 40127, Bologna,
Italy\\
$^{7}$ Instituto de Astrof\'isica de
Andalucia-CSIC, Glorieta de la Astronom\'ia s/n, 18008, Granada, Spain}
\begin{document}

\date{Accepted 2012 May 10. Received 2012 May 10; in original form
2012 May 10}

\pagerange{\pageref{firstpage}--\pageref{lastpage}} \pubyear{2002}

\maketitle

\label{firstpage}

\begin{abstract}
We present the results of a spectroscopic survey performed in the outskirts of
the globular cluster NGC1851 with VIMOS@VLT. The radial
velocities of 107 stars in a region between 12$\arcmin$ and 33$\arcmin$ around the
cluster have been derived. We clearly identify the cluster stellar population 
over the entire field of view, indicating the presence of a significant fraction
of stars outside the tidal radius predicted by King models.
We also find tentative evidence of a cold ($\sigma_{v}\leq 20 ~km~s^{-1}$) peak in the distribution of
velocities at $v_{r}\sim 180~km~s^{-1}$ constituted mainly by Main 
Sequence stars whose location in the color-magnitude diagram is compatible 
with a stream at a similar distance of this cluster. If confirmed, this evidence 
would strongly support the extra-Galactic origin of this feature.
\end{abstract}


\section{Introduction}
\label{intro_sec}

According to the most widely accepted scenario of Galaxy formation many 
globular clusters (GCs) populating the halo of the Milky
Way formed in satellite galaxies accreted in the past by our Galaxy (Searle
\& Zinn 1978). 
Such a hypothesis, originally suggested by the lack of an
abundance gradient in the GCs at Galactocentric distances $>8$ kpc, is also
supported by many pieces of circumstantial evidence: the presence of an age-metallicity 
relation among the "young"
clusters at large distances from the Galactic center (Mar{\'{\i}}n-Franch et al. 2009), 
their peculiar kinematical properties
(large, energetic orbits of high eccentricity), larger core radii and higher
specific frequency of RR Lyrae stars (Mackey \& Gilmore 2004).
The fundamental concept of this picture is also consistent with theoretical 
ideas of the hierarchical formation of structures on galactic scales 
(White \& Rees 1978) and it should therefore hold in other massive galaxies.
The evidence of a
clear correlation between the large coherent streams in the outer halo of M31 
and the position of its GCs seems to confirm this picture (Mackey et al. 2010).
In the typical event of late accretion the satellite is progressively disrupted
by the Galactic tidal strain and the stripped particles
(stars / clusters/ dark matter) continue to move on orbits similar to that of 
the original galaxy, hence forming multiple filamentary wraps
around the parent galaxy (see Law, Johnston \& Majewski 2005).
On the basis of these considerations, the stellar population of the host galaxy
should still be visible in the surroundings of these GCs (van den Berg 2000) as
a compact overdensity of objects in the phase-space distribution. 
Such a direct evidence of association of GCs to confirmed streams have been noticed
for some GC (e.g. Pal 12 and NGC4147 associated to the
Sagittarius dwarf galaxy, Mart{\'{\i}}nez-Delgado et al.
2002; Bellazzini et al. 2003a) while many other candidates have been
proposed to be associated with the Sagittarius galaxy 
(Dinescu et al. 2000; Bellazzini, Ferraro \& Ibata 2003b), the Monoceros ring (Crane et al.
2003; Frinchaboy et al. 2004) and the Canis Major overdensity (Martin et al.
2004a).

An intriguing case is
represented by the GC NGC1851: this cluster is part, together
with NGC1904, NGC2298 and NGC2808, of an apparent system of GCs confined
in a sphere with radius 6 kpc (Bellazzini et al. 2003b) with positions which 
are compatible with the predicted orbital path of the Canis Major stream (Martin et
al. 2004a, Conn et al. 2005). In their discovery paper, Martin et al. (2004a) suggested that the
same episode of accretion which produced 
the Canis Major overdensity was also responsible for the Monoceros ring feature previously observed by
Newberg (2002). N-body simulations by Pe{\~n}arrubia et al. (2005) indicate that the
debris of a satellite with the kinematical properties of Monoceros would
indeed align along the line-of-sight of NGC1851, but the predicted radial
velocity of the stream ($v_{r}\sim 90~km~s^{-1}$) would not be compatible with 
its association with the cluster. However, the nature of these substructures have been questioned
by some authors who claimed that the observed overdensities could be due to the 
Galactic warp (Momany et al. 2004; L{\'o}pez-Corredoira et al. 2006, 2007) and/or flare
(Momany et al. 2006; Hammersley \& L{\'o}pez-Corredoira 2011), and a heated debate
is still ongoing in the scientific community (see also Martin et al. 2004b;
Mart{\'{\i}}nez-Delgado et al. 2005; Vivas \& Zinn 2006; Moitinho et al. 2006; 
Butler et al. 2007; Natarajan \& Sikivie 2007; Carraro et al. 2007, 2008; Conn et al. 2007,
2008, 2012; de Jong et al. 2007; Piatti \& Claria 2008; 
Younger et al. 2008; Kazantzidis et al. 2008; 
Casetti-Dinescu et al. 2006, 2008, 2010; Mateu et al. 2009; Chou et al. 2010; Sollima et al. 2011; 
Michel-Dansac et al. 2011, Meisner et al. 2012).

On the other hand, the accretion origin of NGC1851 has been also suggested by
Carretta et al. (2010) on the basis of the presence of
self-enrichment signatures (the Na-O anticorrelation) in both the two
cluster stellar populations (previously discovered by Milone et al. 2008;
see also Alcaino et al. 1990, Lee et al. 2009).
In particular, several observations (the observed small
metallicity spread, the bimodal distribution of Horizontal Branch and SubGiant 
Branch stars, the different content of neutron-capture elements in the 
metal-rich
and metal-poor components, as well as the Na-O anticorrelation later found
among both blue and red HB stars by Gratton et al. 2012) are better
explained by two originally distinct clusters.
They concluded that NGC1851 could be formed by the merger of two GCs, an
occurrence that is largely unlikely within the Milky
Way but more frequent within dwarf galaxies (van den Bergh 1996).

Furthermore, recent photometric analyses (Olszewski et al. 2009) have also reported that the shape of the
density profile of this cluster deviates from the typical King profile, showing a
power-law decline visible up to 20$\arcmin$ from the cluster center (see also
Carballo-Bello et al. 2012).
Carballo-Bello \& Mart{\'{\i}}nez-Delgado (2010) proposed that part of the
extended stellar population surrounding this stellar system could belong to a 
low-surface brightness stellar stream surrounding this cluster.
Numerical simulations by Bekki \& Yong (2012) showed that the same feature would
be observed if this cluster is the nucleus or a nuclear star cluster formed
within a nucleated dwarf galaxy later accreted by the Milky Way.

In this paper we present the analysis of a sample of 107 spectra of stars
observed in the outskirts of NGC1851 with the aim of studying the distribution
of radial velocities in the region surrounding this stellar system. 
Sect. 2 is devoted to the description of the dataset
and of the reduction procedure. In Sect. 3 the distribution of
velocities is presented and analysed. The estimate of the distance and surface
brightness of an hypothetical satellite possibly revealed in our observations 
is presented in Sect. 4. We discuss our results in Sect. 5.

\section{Observations and Data Reduction} 
\label{obs_sec}

\begin{table*}
 \centering
 \begin{minipage}{140mm}
  \caption{Radial velocities of target stars.}
  \begin{tabular}{@{}lcccccr@{}}
  \hline
   ID & RA (J2000) & Dec (J2000) & B & R & $v_{r}$\footnote{The derived radial 
velocities are available in electronic form at the
CDS (http://cdsweb.u-strasbg.fr/).}     & $\epsilon_{v}$\\
      & deg        & deg         &   &   & $km~s^{-1}$ & $km~s^{-1}$\\
 \hline
   451 &  77.9930363 &  -40.0237255 &  16.573 & 15.313 &   69.4  &   30.0\\
   225 &  78.2332663 &  -39.8095145 &  16.535 & 15.433 &  -71.4  &    8.1\\
     9 &  78.5077249 &  -39.7947895 &  16.758 & 15.754 &   69.4  &   11.0\\
   229 &  78.2059195 &  -39.8067570 &  17.025 & 15.613 &  124.0  &   22.3\\
   109 &  78.3479570 &  -39.6536523 &  17.103 & 15.810 &   99.0  &   18.6\\
   232 &  78.1482069 &  -39.7765154 &  17.167 & 16.114 &  155.8  &   11.2\\
   580 &  78.1494460 &  -39.8404664 &  17.295 & 16.271 &   34.4  &   28.7\\
   346 &  78.0347546 &  -39.7726088 &  17.314 & 16.108 &   79.0  &   14.7\\
   466 &  77.9840459 &  -39.9916386 &  17.337 & 16.312 &  -20.0  &    4.4\\
   579 &  78.1375767 &  -39.9845703 &  17.459 & 16.351 &   -1.4  &   13.9\\

\hline
\end{tabular}
\end{minipage}
\end{table*}

\begin{figure}
 \includegraphics[width=8.7cm]{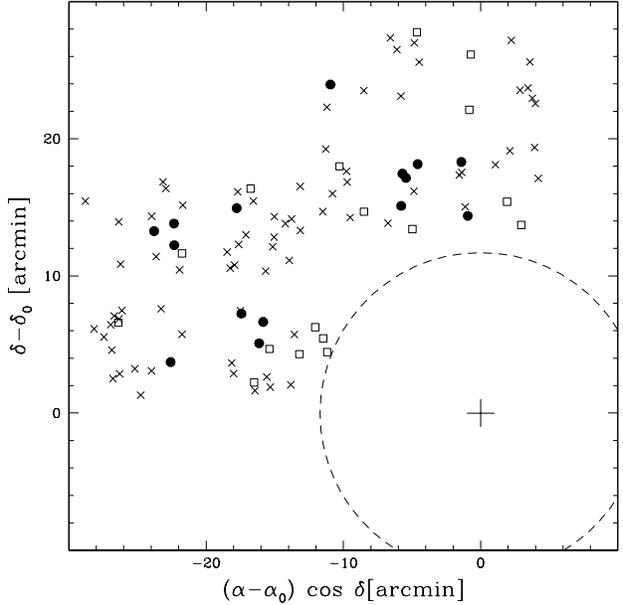}
\caption{Map of the region sampled by the VIMOS observations. North is up,
east towards the right-hand side. The adopted position of the cluster center is
$(\alpha_{0},\delta_{0})=(05h14m06.76s,  -40^{\circ}02\arcmin47.6\arcsec)$ from
the Harris (1996) catalog, 2012 edition.
Stars of the "NGC1851", "stream" and "field" 
samples are marked with
open squares, filled circles and crosses, respectively. 
The cluster center and tidal
radius (from Carballo-Bello et al. 2012) are indicated by the black cross and the
dashed line, respectively.}
\label{map}
\end{figure}

\begin{figure}
 \includegraphics[width=8.7cm]{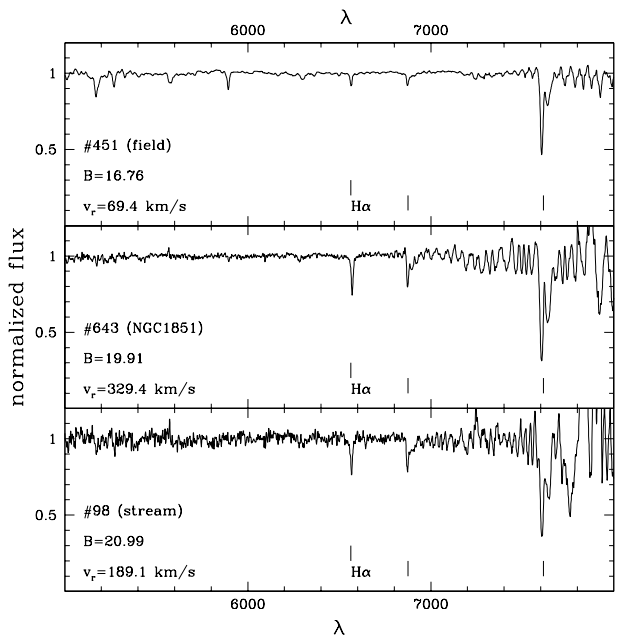}
\caption{Spectra of the field star \#451 (top panel), of the 
NGC1851 star \#643 (middle panel) and of the "stream" star \#98 (bottom panel). 
The H$\alpha$ line and the telluric bands used in the analysis are also indicated.}
\label{sho}
\end{figure}

\begin{figure}
 \includegraphics[width=8.7cm]{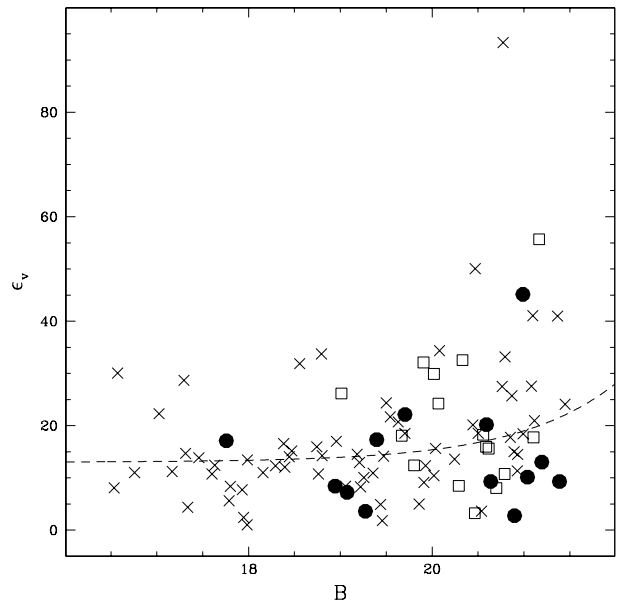}
\caption{Velocity uncertainties of the 96 stars of our sample with at least
three reliable estimates of velocity as a function of 
their B magnitude. Stars of the "NGC1851", "stream" and "field" 
samples are marked with
open squares, filled circles and crosses, respectively. 
The bestfit exponential relation (see
Sect. \ref{obs_sec}) is
shown with a dashed line.}
\label{err}
\end{figure}

Observations have been performed during three nights on October 2008 at the 
Very Large Telescope's (VLT) UT3 at the European Southern Observatory (ESO; 
Cerro Paranal, Chile), equipped with the VIsible MultiObject Spectrograph
(VIMOS). The instrument has been used in Multi-Object Spectroscopy mode with
the medium-resolution (MR) grism coupled with the GG475 filter, allowing a 
spectral coverage from 5000 to 8000 \AA~ with a resolution R$\sim$580.
Masks have been constructed by placing $\sim$110 1.0$\arcsec$-wide slits on target stars selected
from the photometry of Carballo-Bello et al. (2012) sampling a wide range in
magnitude and color ($16<B<22$, $0.6<B-R<2.1$) oversampling the Main Sequence-Turn off
(MS-TO) region where both extratidal stars and surrounding streams are expected
to be located (Mart{\'{\i}}nez-Delgado et al. 2004). We observed two fields centered outside the nominal tidal
radius of the cluster ($r_{t}=11.7 \arcmin$; Carballo-Bello et al. 2012), 
sampling a region between $12\arcmin<d<33\arcmin$ ($r_{t}<d<3~r_{t}$) 
to the cluster center (see Figure \ref{map}).
The spectra have been obtained combining four 1955 s long exposures secured in good 
seeing conditions (FWHM $< 1.0 \arcsec$), reaching a signal-to-noise
ratio $10<S/N<300~px^{-1}$ depending on the target magnitude.
The one-dimensional spectra have been extracted with the VIMOS pipeline.
Unfortunately, the solution in wavelength calibration provided by the pipeline
was not satisfactory, mainly because of the misalignment of targets within the
slits (see also Giuffrida et al. 2010). To overcome this problem, an additive 
shift in wavelength\footnote{The choice of the additive shift is based on the
fact that the misalignment within the aperture produces a constant shift in pixel
when passing through a grism.} has been applied to match the strong telluric absorption 
lines at $6875$ \AA~ and $\sim7615$ \AA.
The spectra were then continuum-normalized with IRAF. 
Spectra of three program stars are shown in Figure \ref{sho} 
to illustrate the quality of our data.

Radial velocities have been then obtained by cross-correlating the spectra of
the individual exposures with a
GIRAFFE solar spectrum\footnote{http://www.eso.org/observing/dfo/quality/GIRAFFE/pipeline/solar.html} 
smoothed to the resolution of our targets. For the cross-correlation we used the
region of the H$\alpha$ line (6540 \AA~ $<\lambda<6590$ \AA) which is well visible
even in the spectra of the faintest stars of our sample. Velocities have been 
corrected for heliocentric velocity and averaged\footnote{The derived radial 
velocities are available in electronic form at the
CDS (http://cdsweb.u-strasbg.fr/).}. Errors have been calculated as the r.m.s of
repeated measures for the 96 stars with at least three reliable estimates (see Fig. \ref{err}). 
These stars have been used to fit an exponential relation as a function of the B 
magnitude that has been used to assign the errors to the other stars. 
The final dataset consists of 107
radial velocities with average uncertainties of $\sim15~km~s^{-1}$. 
The radial velocities of
the entire sample together with their coordinates and magnitudes are listed in Table 1.
Line strength indices of the Mg triplet at $\sim5175$ \AA~ and the Fe line at
$\sim5265$ \AA~ have been also measured adopting the bandpasses defined by 
Worthey (1994). However, given the low resolution of our spectra, the derived
metallicities have large uncertainties which prevent their use to identify interlopers
from the Galactic halo and thick disk.

\section{Results}
\label{res_sec}

\begin{figure}
 \includegraphics[width=8.7cm]{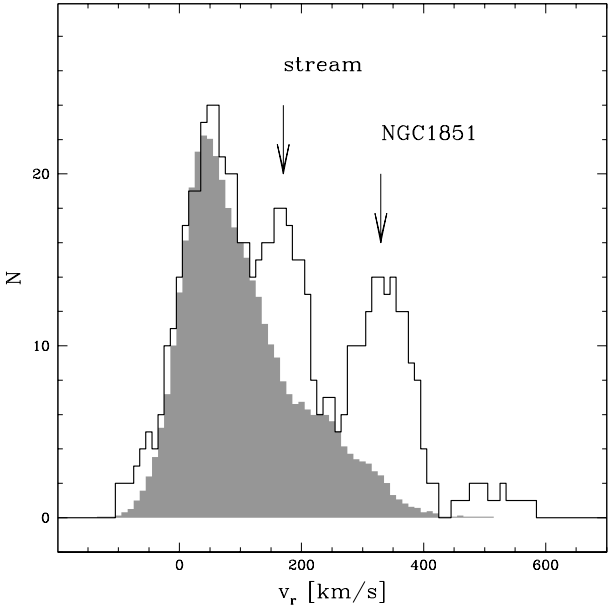}
\caption{Velocity distribution of the 107 stars of our sample (empty histograms). 
The prediction of
the Galactic model of Robin et al. (2003) is overplotted as grey histograms.
The location of the "stream" and of the "NGC1851" peak are indicated with
arrows.}
\label{distr}
\end{figure}

\begin{figure}
 \includegraphics[width=8.7cm]{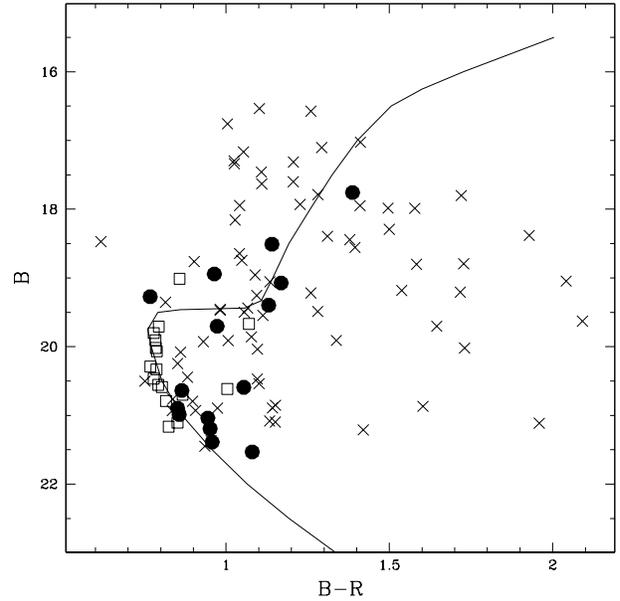}
\caption{B vs. B-R CMD of the target stars. Stars of the "NGC1851", "stream" and "field" 
samples are marked with
open squares, filled circles and crosses, respectively. 
The mean ridge line of NGC1851 is
overplotted.}
\label{cmd}
\end{figure}

\begin{figure}
 \includegraphics[width=8.7cm]{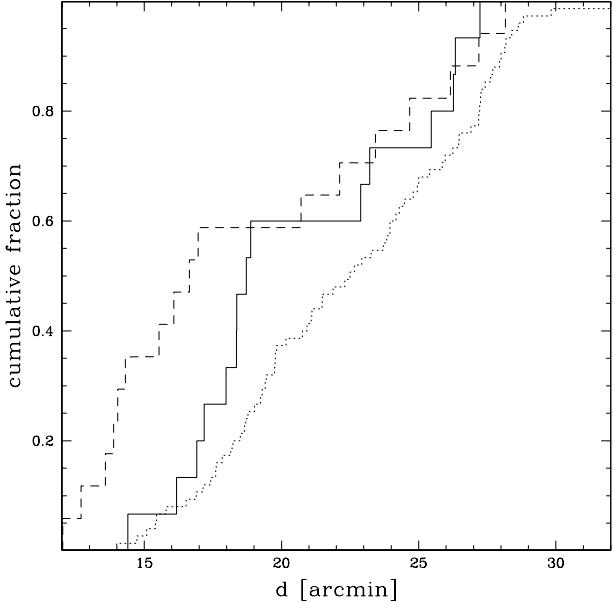}
\caption{Cumulative radial distribution of the "stream" sample (solid line), 
"NGC1851" sample (dashed line) and "field" sample (dotted line) as a function 
of the distance from the cluster center.}
\label{ks}
\end{figure}

In Figure \ref{distr} the distribution of radial velocities of the entire sample
is shown. The histogram has been constructed using the naive estimator
(Silvermann 1986)\footnote{The naive estimator is defined as a series of
histograms having a given width (the "window width") defined over a grid of 
sampling points separated by a given amount (the "step"). At odds with ordinary
histograms, in the naive estimator the step is not necessary equal 
to the window width.
Thus, every point of the grid is 
the centre of a sampling interval, freeing the histogram from a particular choice 
of bin positions. The 
choice of bin width controls the amount by which the data are smoothed to 
produce the estimate.} with a window width of 60 $km~s^{-1}$ and a step of 10 $km
s^{-1}$. The window width has been chosen using the bootstrap-based 
algorithm described by Faraway \& Juhn (1990). 
The distribution shows three well separated peaks: {\it i)} a prominent peak
at $v_{r}\sim 30~km~s^{-1}$ due to the foreground population of disk stars, 
{\it ii)} an overdensity of stars at $v_{r}\sim 180~km~s^{-1}$ (hereafter referred as
"stream peak") and
{\it iii)} a peak at $v_{r}\sim330~km~s^{-1}$ correspondent to the bulk velocity of 
NGC1851 ($320.3\pm0.4~km~s^{-1}$, Carretta et al. 2011; "NGC1851 peak").
The prediction of the Galactic model of Robin et al. (2003) is overplotted.
As the target stars have not been selected in an unbiased way (see Sect.
\ref{obs_sec}) the comparison with the Galactic model has been performed
adopting the
following procedure: we retrieved 100 synthetic field catalogs from the
Besan\c{c}on
website\footnote{http://model.obs-besancon.fr/} each of them covering 1 sq.
deg. in the direction of NGC1851. For each catalog we randomly associated a 
synthetic object to each target according to their euclidean distance in the B vs.
R plane. A random gaussian distributed shift with dispersion equal to the
velocity error of each target has been then added to each synthetic object. 
The procedure is repeated for all the 100
extractions and the naive estimator has been calculated on the overall
catalog. The distribution has been then normalized to the observed one requiring
the same number of stars at velocities $v_{r}<120~km~s^{-1}$ (where only field stars
are expected). 
It is apparent that both the stream and the NGC1851 peaks
are not reproduced by the Galactic model of Robin et al. (2003).
The peaks shown in Fig. \ref{distr} remain apparent even when
different choices of the bin width are made ($40<\Delta v_{r}/km~s^{-1}<100$).
In the following analysis we defined two samples of stars in the radial velocity range
$160<v_{r}/km~s^{-1}<210$ (encompassing the "stream peak") and
$310<v_{r}/km~s^{-1}<430$ (around the "NGC1851 peak"), and a "field sample"
containing the remaining stars. They contain 15, 17 and 75 stars, respectively. 
The comparison with the
Galactic model indicates an expected contamination from Galactic interlopers of
$\sim6.8\pm0.6$ stars in the "stream" sample and $\sim2.5\pm0.3$ stars in the
"NGC1851" one. So, assuming Poisson fluctuations in number counts, the
statistical significances of the "stream" and "NGC1851" peaks are 2.2 and 
3.4 $\sigma$, respectively. The above significances, in particular that of the 
"stream peak", are slightly sensitive to
the normalization of the Galactic model to the observed data: in the extreme
assumption of a normalization to the number of targets over the entire velocity range (instead of the
above quoted range $v_{r}<120~km~s^{-1}$) the significances of the "stream" and
"NGC1851" peaks would decrease to
1.6 and 3.3 sigma, respectively.
We then performed a Kolmogorov-Smirnov test (independent on the normalization of the
two samples): the probability that the observed velocity distribution and that
predicted by the Robin et al. (2003) model are drawn from the same population
turns out to be 0.003\%. If we exclude the stars of the "stream peak" the
probability decreases below $<0.001\%$, while excluding the stars of the
"NGC1851" peak we obtain a probability of 3.6\%.

The observed velocity dispersion of the two peaks are
$\sigma_{v,stream}=13^{+5}_{-4}~km~s^{-1}$ 
and $\sigma_{v,NGC1851}=23^{+10}_{-4}~km~s^{-1}$ i.e.
comparable with the observational uncertainty. This suggests that these features
are kinematically cold (with an intrinsic velocity dispersion 
$\sigma_{v}\leq 20 ~km~s^{-1}$).

In Figure \ref{cmd} the location of the stars of the "stream peak" and
of the "NGC1851 peak" in the B, B-R color-magnitude diagram (CMD) are shown.
It can be noted that the stars of the "NGC1851 peak"
nicely follow the mean ridge line of the cluster populating the cluster MS
region.
The distribution of the "stream peak" stars is instead more scattered covering
the entire range of magnitudes of the sample while being
confined in a narrow range in color ($0.8<B-R<1.4$), partly overlapping the
red side of the MS of NGC1851.

Fig. \ref{ks} shows the cumulative radial distribution of the three groups of stars
(field, stream and NGC1851) as a function of the distance from the cluster center. 
Note that the stars of the "NGC1851"
sample are more concentrated toward the cluster center with respect to those of
the other groups: a Kolmogorov-Smirnov test indicates a probability of less than
0.2\% that the field and NGC1851 samples are extracted from the same population.
The same result can be deduced by splitting the sample in two subsets
according to their distance from the cluster center ($d \lessgtr 20\arcmin$) and calculating the
double-normalized ratio between the number of objects in the "NGC1851" and
"field" samples contained in the inner and outer subsamples
$$R=\frac{N_{1851}^{out}~N_{field}^{in}}{N_{field}^{out}~N_{1851}^{in}}=0.42\pm0.23$$
Note that the bestfit of the density profile provided by McLaughlin \& van der 
Marel (2005) using 
a Wilson (1975) model and that provided by Carballo-Bello et al. (2012)
with the power-law empirical relation by Elson, Fall \& Freeman (1987) 
predict $R=0.05$ and $R=0.24$ in the same radial
range, respectively. So, in spite of the large errors involved, there seems to 
be an overabundace of cluster members beyond 20$\arcmin$ from the cluster center
with respect to the model predictions.
Unfortunately, the small number of objects prevents from any conclusion on the
radial distribution of the "stream" sample.
Also, no clear signatures of azimuthal variation of velocities have been found
in the three samples. Note that the velocity variation of a cold stream over
the field of view covered by our observations should not exceed few $km~s^{-1}$
(Pe{\~n}arrubia et al. 2005) so, because of the relatively large velocity 
uncertainties and the likely presence of field interlopers, we do not expect to
detect such differences.

\section{The stream hypothesis}

\begin{figure}
 \includegraphics[width=8.7cm]{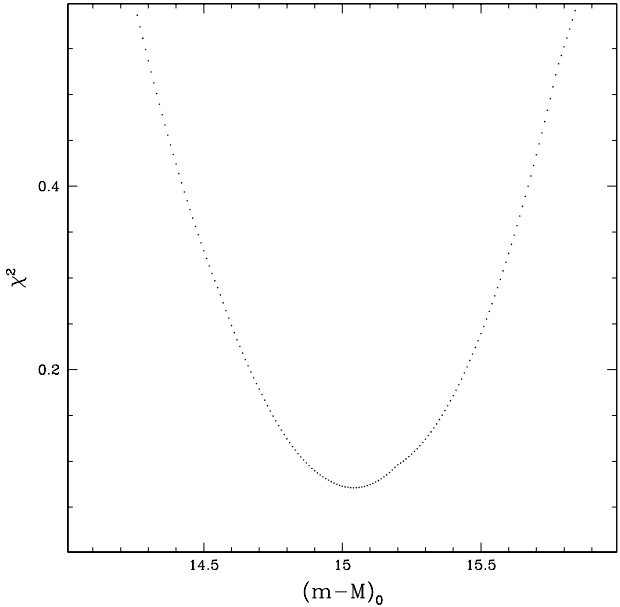}
\caption{$\chi^{2}$ of the isochrone fitting of the "stream" sample as a function 
of the adopted distance modulus for the hypothetical stream population.}
\label{dist}
\end{figure}

\begin{figure}
 \includegraphics[width=8.7cm]{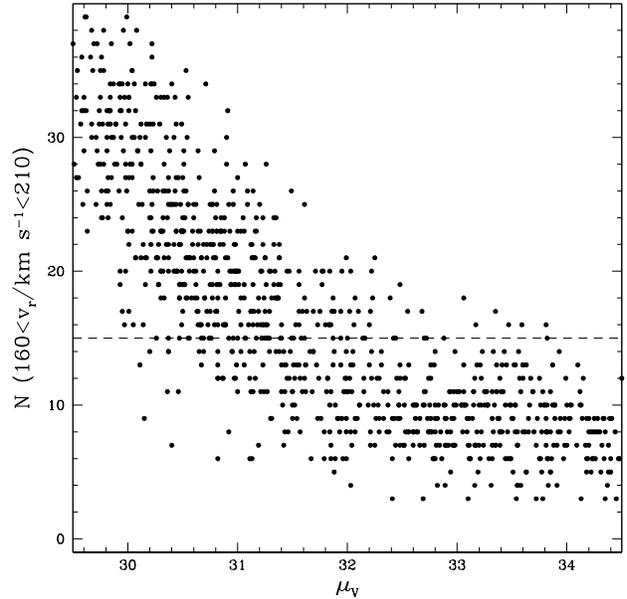}
\caption{Number of stars in the "stream" sample velocity range ($160<v_{r}/km
s^{-1}<210$) as a function of the satellite surface brightness for the set of synthetic
stream population. The observed value is indicated by the dashed line.}
\label{sb}
\end{figure}

While the natural interpretation of the "NGC1851" peak in the velocity
distribution is that these stars are cluster members, 
the origin of the "stream" peak is not clear. An intriguing possibility is that
it is a debris of an accreted satellite in the direction of NGC1851.
In this section we try to estimate the distance and the surface brightness of
this feature, in the hypothesis that it is constituted by the debris of the
Monoceros ring.

For this purpose, we simulated a synthetic CMD
of the satellite by randomly extracting a population of stars from a Kroupa 
(2001) mass function and interpolating through a set of Marigo et al. 
(2008) isochrones with an age of 9.2 Gyr (Sollima et al. 2011), a metallicity of
[Fe/H]=-0.95 with a spread of $\sigma_{[Fe/H]}=0.15$
(Ivezi{\'c} et al. 2008; Meisner et al. 2012) and a reddening of E(B-V)=0.04 (Schlegel, Finkbeiner \&
Davis 1998).

To estimate the distance of the stream we performed an iterative weighted fit of 
the location of the "stream" sample stars with the same isochrone used to
compute the synthetic CMD.
As a first step we chose a first guess of the distance
modulus and converted absolute magnitudes to apparent ones.
For each star a weight has been then calculated as the ratio between the
densities of stars at the target (B, B-R) position in the synthetic
satellite CMD and in the Robin et al. (2003) Galactic model CMD.
Only stars in the "stream sample" velocity range have been used to calculate the
Galactic field density.
The derived bestfit distance has been then adopted for the next iteration and
the procedure is repeated until convergence. The above algorithm converges after
few iterations and appears to be insensitive to the first guess of the distance.
The $\chi^{2}$ of the fit as a
function of the distance modulus is shown in Fig. \ref{dist}. The bestfit
distance modulus turns out to be $(m-M)_{0}=15.04\pm0.38$ corresponding to a
distance of $10.2\pm1.8$ kpc.

To estimate the surface brightness of this object we simulated the synthetic CMD
of the satellite adopting the above derived distance and left the number of 
simulated objects as a free parameter.
A velocity extracted from a gaussian distribution with 
$<v_{r}>=180~km~s^{-1}$ and $\sigma_{v}=10~km~s^{-1}$ has been assigned to each star 
of the satellite. 
The CMD of the Galactic model of Robin et al. (2003) has been then added and 
a synthetic object has been associated to each target of the entire
sample\footnote{At odds with the procedure described in Sect. \ref{res_sec} we
excluded here the NGC1851 stars with $v_{r}>310~km~s^{-1}$ as they are not included
in our simulation.} (see
Sect. \ref{res_sec}). A random gaussian distributed shift with dispersion equal to the
velocity error of each target has been then added to the synthetic
velocities to mimic the observational errors. 
The number of objects falling in the velocity range of the 
"stream sample" has been then counted and the surface brightness of the
satellite has been estimated by summing the V fluxes of all the simulated
sources.
The procedure has been repeated 100 times for each adopted number of satellite 
object and the distribution in the
$\mu_{V}-N_{stream}$ plane is shown in Fig. \ref{sb}. 
As expected, the brighter is the adopted satellite the larger is the number of
stars in the "stream" range of velocity, while at surface brightness lower than 
$\mu_{V}\leq 32.8$ the number of objects in the interesting velocity range
asymptotically tends to the number of halo stars. The mean surface brightness to observe
15 object turns out to be $\mu_{V}=31.4\pm0.7~mag~arcsec^{-2}$.

\section{Discussion}

\begin{figure}
 \includegraphics[width=8.7cm]{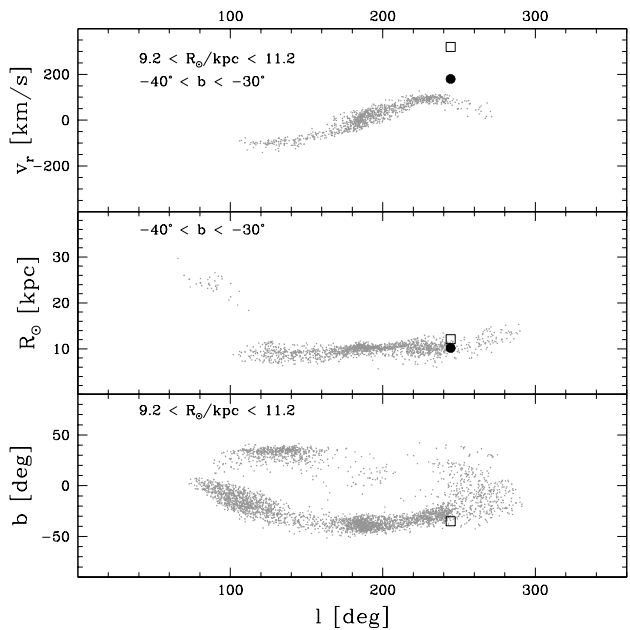}
\caption{Orbital path of the Monoceros stream predicted by the Pe{\~n}arrubia et
al. (2005) model. The bottom, central and upper panels show the distribution of
stars in Galactic longitude, heliocentric distance and radial velocity,
respectively, as a function of the Galactic latitude. Only particles in the
ranges indicated in each panel have been plotted.
The position of NGC1851
is marked by the open square. In the upper and central panels the position of 
the "stream" sample is marked by the filled dot.}
\label{mon}
\end{figure}

We report the results of a survey of radial velocities in the outskirts of the
GC NGC1851. We clearly detect the signal of the cluster stellar population as a
strong overdensity of objects with a velocity and location in the CMD compatible
with that of the cluster MS, and more concentrated with respect to the Galactic field
population. Such a detection confirms that a significant number of cluster 
stars are present up to 25$\arcmin$ from the cluster center, well beyond the
King tidal radius, as already reported
by Olszewski et al. (2009) and Carballo-Bello et al. (2012). It is worth
noting that the truncation radius predicted by dynamical models depends on their
adopted functional form of the distribution function, a choice which is somehow
arbitrary. For instance, McLaughlin \& van der Marel (2005) showed that Wilson
(1975) models, predicting larger tidal radii with respect to King (1966) ones,
better fit the overall shape of many GCs. For NGC1851 the tidal radius estimated
by McLaughlin \& van der Marel (2005) using Wilson
(1975) models is 44.7$\arcmin$, well beyond the extent of our dataset. 
Anisotropy and mass segregation can also play a role in shaping the radial density profile of the
cluster (Gunn \& Griffin 1979). However,
a physical (model-independent) limit to the distance of bound stars is given by
the Jacobi radius at which equilibrium between the effective potentials of the 
cluster and of the host galaxy settles. Allen, Moreno \& Pichardo
(2006) estimated a Jacobi radius of $\sim17\arcmin$ for NGC1851. Thus, the outermost targets
observed here could be unbound objects. Also in this case, however, caution must
be taken since the estimate of the Jacobi radius depends on the adopted cluster 
orbital parameters and on the Galactic potential, which are subject to
significant uncertainties. Even if all the detected cluster stars would be 
comprised within the Jacobi radius the relative fraction of object at
$d>20\arcmin$ exceeds the prediction of all the commonly used dynamical models.
The overabundance of these object can be explained assuming these stars to be "potential escapers" 
i.e. stars whose orbital energy already exceeded the potential of the
Lagrangian point but whose orbits have still not 
intersected it (K{\"u}pper et al. 2010). In this scenario, these stars are
expected to have also a velocity dispersion significantly larger than those
predicted by dynamical models. Unfortunately, the
small sample size and the relatively large uncertainties of our data do not
allow to confirm this hypothesis. Another possibility is that these objects have
been heated by the relaxation process after the last cluster orbital 
pericentric passage, moving over the tidal boundary (Baumgardt \&
Makino 2003).

We also found a peak at $v_{r}\sim180~km~s^{-1}$ not predicted by the Galactic
model of Robin et al. (2003) corresponding to a
cold ($\sigma_{v}\leq 20~km~s^{-1}$) population of stars whose location in the CMD is
compatible with a distance of $\sim$10 kpc and a surface brightness of 
$\mu_{V}=31.4\pm0.7~mag~arcsec^{-2}$. Note that the deepest wide-field photometric 
analyses of this cluster reached a surface brightness limit of $\mu_{V}\leq30$
(Olszewski et al. 2009; Carballo-Bello et al. 2012) and could not detect such a feature.
It is necessary to point out that the significance of such a peak is 2.2 $\sigma$,
leaving a 1.5\% probability of false detection (3.6\% according to the
Kolmogorov-Smirnov test). A non-optimal wavelength
calibration of the spectra can also introduce some artifact in the velocity
distribution, although it is not expected to produce a spurious peak.
Moreover, it is also possible that the Galactic model of Robin et al. (2003)
underpredicts the fraction of stars in the velocity interval around 180 $km
s^{-1}$
(constituted almost exclusively by halo stars) by a factor of two. 
However, the mean velocity and the relatively high Galactic
latitude ($b=-35.03^{\circ}$) of this object makes impossible a connection with the
warped/flared disk (see Sect. \ref{intro_sec}).
If confirmed this evidence
would constitute a strong indication of the presence of a stream in the
direction of NGC1851. 
An identification of this feature with known streams is not easy because of the
large uncertainties in the orbit of these satellites. The prediction
for the Monoceros stream of the N-body model by Pe{\~n}arrubia et al. (2005) 
indicates that a high latitude debris of this stream should align with NGC1851 at
a distance compatible with the range estimated here (including the NGC1851, see
Fig. \ref{mon}). 
However, the connection
of this cluster with the stream is unlikely because of the large velocity
difference between these two objects ($\Delta v>200 ~km~s^{-1}$). The velocity difference between the
"stream" sample identified here and the prediction for the Monoceros ($\Delta
v\sim90 ~km~s^{-1}$) seems to exclude also a connection between them. Note however that our
observations are located far away from the kinematical detection used by
Pe{\~n}arrubia et al. (2005) to constrain the Monoceros orbit, so many
uncertainties (including the adopted Galactic potential) make this comparison
uncertain.
Unfortunately, the small sample size and the low resolution of our data
prevent us a statistically significant detection. Further high-resolution spectroscopic
studies over a larger field of view are necessary to confirm this evidence.

\section*{Acknowledgments}

A.S. acknowledge the support of INAF through the 2010
postdoctoral fellowship grant. RG, EC, AB and SL acknowledge
the PRIN INAF "Formation and Early Evolution of Massive Star Clusters". 
JP acknowledges support from the Ram\'on y Cajal Program
as well as by the Spanish grant AYA2010-17631 awarded by the Ministerio of
Econom\'ia y Competitividad. We thank Michael West, the referee of our paper,
for his helpful comments and suggestions.


\label{lastpage}

\end{document}